# Enhanced Authentication and Locality Aided - Destination Mobility in Dynamic Routing Potocol for MANET

Sudhakar Sengan[1]
Lecturer, Department of CSE,
Nandha College of Technology ,
Erode -TamilNadu – India
sudhasengan@gmail.com

Dr.S.Chenthur Pandian[2]
Principal
Selvam College of  Technology
Namakkal -TamilNadu – India
chenthur@rediffmail.com

*Abstract* — Mobile Ad Hoc Network (MANET) is an emerging area of research in the communication network world. As the MANET is infrastructure less, it is having dynamic nature of arbitrary network topology. So, it needs set of new networking strategies to be implemented in order to provide efficient end to end communication. Node activities such as sending or receiving data is highly traceable. Nodes are vulnerable to attacks and disruptions. To identify such nodes, a method of direct validation is proposed. Since it is unlikely for 2 ad hoc nodes to stay at the same position concurrently, the match between a position and ID is definitely unique.This information is obtained via global positioning system (GPS) and location services. In the routing protocol, location information is distributed between nodes by means of position beacons. Routing schemes rely on the cooperation and information exchanged among the nodes. Here in addition to node ID, extra information such as positions of the nodes is used for making routing decisions. Its neighbouring nodes receive the request and content to access the channel for becoming the next hop using Receiver Contention Channel Access Mechanism. A receiver that is geographically closer to the destination is assigned a higher priority and can win the contention. The destination also finds the corresponding authentication code according to the position carried in the rreq and encrypts the code with the secret key of its secret key pair.The encrypted result is included in the rrep and sent to the source.The source finds out whether it reaches the right destination by decrypting the information with the destination's key and comparing the authentication code with the one it obtained through the position request.

To avoid intruder for routing, Packet Dropping, WatchDog, SYBIL Attacks and PathSelector are used.The watchdog identifies misbehaving nodes, while the Pathselector avoids routing packets through these nodes. The watchdog, the path selector is run by each server. Each Server maintains a rating for every other node it knows about in the VHR. In our proposed model, the route selection is a function of following parameters: hop count, trust level of node and security level of application. In this paper, to focus on secure neighbor detection, trust factor evaluation, operational mode, route discovery and route selection. The paper mainly address the security of geographic routing.The watchdog identifies misbehaving nodes, while the Pathselector avoids routing packets through these nodes. The watchdog, the pathselector is run by each server. In order to keep the source informed about the destination's mobility, the destination keeps sending the alert message to its previous hop telling that it has changed its position and any reference to it for data packet forwarding be informed to the VHR server.

I. INTRODUCTION

Wireless networking is an emerging technology that allows users to access information and services electronically, regardless of their geographic position. Wireless networks can be classified in two types:

A.  *Infrastructure networks:*

Infrastructure network consists of a network with fixed and wired gateways. A mobile host communicates with a bridge in the network (called base station) within its communication radius. The mobile unit can move geographically while it is communicating. When it goes out of range of one base station, it connects with new base station and starts communicating through it. This is called handoff. In this approach the base stations are fixed.

B.  *Infrastructureless (Ad hoc) networks:*

In ad hoc networks all nodes are mobile and can be connected dynamically in an arbitrary manner. All nodes of these networks behave as routers and take part in discovery and maintenance of routes to other nodes in the network. Ad hoc networks are very useful in emergency search-and-rescue operations, meetings or conventions in which persons wish to quickly share information, and data acquisition operations in inhospitable terrain. These ad-hoc routing protocols can be divided into two categories:

C.  *Table-driven routing protocols:*

In table driven routing protocols, consistent and up-to-date routing information to all nodes is maintained at each node.

*D. On-Demand routing protocols:*

In On-Demand routing protocols, the routes are created as and when required. When a source wants to send to a destination, it invokes the route discovery mechanisms to find the path to the destination.

II. ROUTING PROPOSAL

The wireless networks with infrastructure support, a base station always reaches all mobile nodes, this is not the case in an ad hoc network. Thus, routing is needed to find a path between source and destination and to forward packets appropriately. In Traditional Routing Algorithms like AODV[1], DSR[2], DSDV[3],a node has to disclose its ID in the network for building a router.Node activities such as sending or receiving data is highly traceable. Nodes are vulnerable to attacks and disruptions. Routing schemes rely on the cooperation and information exchanged among the nodes.

These routing algorithms that rely largely or completely on location information (based on position). Here in addition to node ID, extra information such as positions of the nodes is used for making routing decisions. Since it is unlikely for 2 ad hoc nodes to stay at the same position concurrently, the match between a position and ID is definitely unique.Hence, in these algorithms , when positions are revealed for routing , there is no need of node IDs. Hence node anonymity can be maintained.However such algorithms rely on position exchange among the neighbouring nodes. Such time based position exchange messages make a node highly traceable. The trajectory of a node movement can be well known to other nodes even when its node ID is intentionally hidden. Hence there is lack of privacy in traditional position based ad hoc routing algorithms.

The destination's position alone is revealed for routing purposes thereby maintaining the privacy of other nodes IDs. For routing discovery, a node sends out a routing request. Its neighbouring nodes receive the request and contend to access the channel [5]. for becoming the next hop using Receiver Contention Channel Access Mechanism. A receiver that is geographically closer to the destination is assigned a higher priority and can win the contention. Once the route is built, only Pseudo IDs are generated and are used by the nodes participating in the route. Nodes that get the access to the channel by winning the contention may maliciously drop packets. To avoid such nodes from becoming part of the route, Packet Dropping [6], WatchDog, SYBIL Attacks and PathSelector are used. Certain nodes may try to win the contention by reporting false Pseudo IDs as their own ID. To identify such nodes, a method of direct validation is proposed.

III. POSITION MANAGEMENT

Virtual Home Region (VHR) based distributed secure position service system. An Ad Hoc node is assumed to be able to obtain its own geographic position [7]. It is assumed that a source is able to get the position of its destination. Each node has a geographical region around a fixed center called the Virtual Home Region(VHR).The relationship between a node ID and the VHR follows a hash function that is predefined and known to all the nodes who join the network.

A number of servers which are also ad hoc nodes are distributed in the network. A node updates its position to the servers located in its VHR to which other nodes send position requests acquiring this node's position. Only a small number of trusted nodes can act as position servers. A node updates its position to its VHR when the distance between its current position and the last reported position exceed a threshold value. When the source gets the position of its destination, it also gets the time when the position is updated and an authentication code. The time is needed for accuracy and the code can be any random number generated and sent to the position server by the destination.

*A.Position Verification*

The location based routing protocol require that a node be able to identify it's own position and position of destination node. This information is obtained via global positioning system (GPS) [8] and location services. In the routing protocol, location information is distributed between nodes by means of position beacons.

All network used in MANETs have a maximum communication range. Based on this properties, we define acceptance range threshold 'T'. Position beacons received from nodes that are at position larger than 'T' away from current position of receiving nodes can be discarded. Position can also be verified based on the mobility of the node. It is assumed that all nodes move at well defined speed. When receiving a beacon the node records the arrival time of beacon. On receiving subsequent beacons, the node checks the average speed of nodes between two position in two beacons. If the average speed exceeds mobility grade T, the position beacon is discarded.

```
A  receives beacon from B
if distance(A's position, B's position) = T
    if B is in A's neighbor table
        update the position information of B
    else
        add B's ID, position details in  A's table
    else
        reduce trust value of B drop  beacon
```

Algorithm for position verification based on transmission range.

```
A receives beacon from B  , t=time of last beacon
                from B
if B is not in A's neighbor table add B's ID, position
              details in A's table   else
        old=position of B in A's table
        new=position information in beacon
        speed=distance(new,old)/(current time-t)
            if speed=Max.speed
                update position and time  details
            else
                reduce trust level of B drop beacon
```

Algorithm for position verification based on mobility

## IV. ROUTING DISCOVERY

Here, a source discovers its route through the delivery of a routing request to its destination. To find the route to its destination, the source first generates a Pseudo ID for itself through a globally defined Hash Function using its position and current time as its inputs. This procedure makes the probability of 2 active nodes having the same pseudo ID negligible. The source then sends out a Routing Request(*rreq*) message that carries Position of the destination, distance from this source to the destination and the source Pseudo ID.

The neighbouring nodes around the source called receivers will receive rreq. A receiver checks to find out whether it is the intended destination. If not, it uses the hash function to generate its own Pseudo ID. The receivers then contend for the wireless channel to send out the hop reply-hrep message. This contention mechanism called the hrep Contention Mechanism is discussed very soon. The receiver who has successfully sent out the hrep will be the next hop. Its pseudo ID is carried in the hrep. On receiving the hrep, the source replies with a confirm message(*cnfm*). Its next hop replies to this message with an ack. On receiving this ack, the source saves the pseudo ID in its routing table.

On receiving the cnfm message, the next hop receiver becomes a sender. The searching for the next hop is continued until the destination receives the rreq message. Finally, the destination sends out a Routing reply(*rrep*) message through the reverse path to the source. The destination also finds the corresponding authentication code according to the position carried in the rreq and encrypts the code with the secret key of its secret key pair. The encrypted result is included in the rrep and sent to the source. The source finds out whether it reaches the right destination by decrypting the information with the destination's key and comparing the authentication code with the one it obtained through the position request.

*Message Flow in routing discovery*

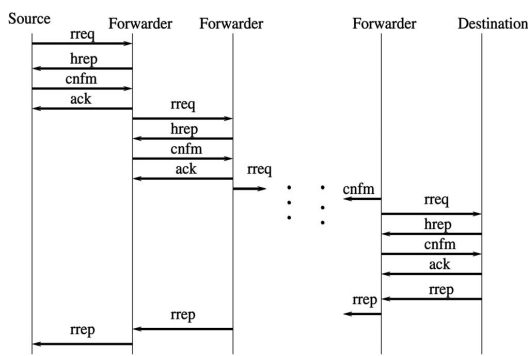

## V. RECEIVER CLASSIFICATION

A receiver determines its node class by finding that, if it is the next hop, how much closer ( this geographic distance is defined as Δd) it can move a rreq from the sender toward the destination. Δd can be calculated because the distance between the receiver and the destination is known based on their positions and the distance between the sender and the destination is carried in the <u>*rreq*</u>.

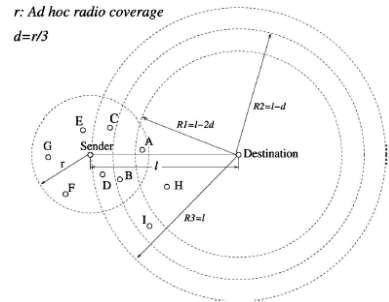

In this example, all nodes except the destination are divided into four node classes. A distance of d is calculated as d = r/3, where r is the maximum radio coverage of the ad hoc channel. Nodes with Δd>2d (e.g. node A, as if falls in the circle centered at the destination with a radius of l-2d belong to class 1, which has the highest priority. Nodes with d ≤Δd<2d(e.g. node B) and nodes with 0≤Δd<d (e.g. nodes C and B) belong to Class 2 and 3 , respectively , and have lower priorities. For nodes E,F,and G, Δd<0. They belong to class 4 and will lead the rreq away from the destination. Other nodes, such as H and I, are out of the sender's transmission range and cannot receive the rreq. Note that the destination is a special node. It has the highest priority to access the channel with a class of 0. In this paper, we investigated the algorithm in which only nodes of class 1, 2, and 3 will contend to be legitimate receivers. A node of class 4 will not attend the contention because it leads a rreq away from the destination. The node classification scheme is used only for simplicity of presentation and will be used in the rest of the paper. In more complicated schemes, rules for node classification can be adaptive based on node density. When the density is high, only the nodes that can greatly reduce the distance between the rreq and the destination should be assigned to the class with a high priority. On the other hand, if the nodes are sparsely distributed, a node which leads the rreq away from the destination can also be a possible legitimate receiver. Such a rule adaptation, for example, can be made by adjusting the value of d. Besides the distance to the destination, other criteria, such as signal quality, the remaining power of a node, and node mobility, can also be considered in node classification.

## VI. RECEIVER CONTENTION CHANNEL ACCESS MECHANISM - (*hrep*)

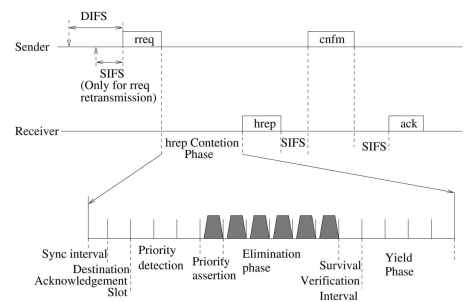

The receiver contention mechanism used in hrep contention phase is EY-NPMA[9] (Elimination Yield Non

Preemptive Priority Multiple Access).The receivers are classified based on how much closer(geographical distance) they can move the rreq from the sender toward the destination. Based on this distance, priorities are assigned to the different nodes.

*A.The hrep contention phase is divided into 3 phases:*
*a.Prioritization Phase :*

This phase allows only the receivers with highest channel access priority among the contending nodes to participate in the next phase. A number of slots, the same as the number of different priority classes are available in this phase. A receiver of priority 3 can send a burst slot 3 only if no burst is sent in the previous 2 slots. This means it has the highest priority. It therefore enters the next phase. If a receiver senses a burst in one of the previous slots, it will quit from hrep contention, cannot enter the next phase and therefore drop rreq.

*b.Elimination Phase:*

This phase starts immediately after the transmission of the prioritization bursts and consists of a number of slots.A receiver in this phase will transmit burst in a randomly selected slot. The receiver transmitting the longest series of bursts will survive. After the end of burst transmission, each receiver senses the channel for the duration of the elimination survival verification slot. If the channel is sensed to be idle, the receiver is admitted to the next phase. Otherwise, it drops itself from contention.

*c.Yield Phase:*

In this phase, a receiver will yield for a number of slots and listens to the channel and if the channel is sensed idle, it sends out a hrep. Otherwise, the receiver loses contention and drops the rreq. When more than one receiver sends out a hrep at the same time, a hrep collision occurs. The sender will have to resend the rreq in such cases.

The next hop is determined by node contention mechanism as illustrated above. A malicious node can always use this most aggressive contention mechanism to become the next hop. Once it is included in a route, it can conduct different attacks such as Packet dropping and false misbehaviour.

*B.Secure Neighbor Detection*

A node N broadcasts[10] a hello message M1 with it's certificate. The target node receiving the message M1 decrypt N's certificate to verify and obtain N's public key. The target node sent the reply through message M2. After receiving the response, N stores the nodes public key and recent location coordinates of the target node in it's neighbor table. Node N records the sending time of M 1 at t0 and receiving time of M2 at t1.

Total delay d = t1 – t0

Distance between the nodes must be less than (d/2) * c, where c is the speed of light. Thus node N can check that the other party is within it's transmission range.

## VII.PACKET DROPPING

When nodes act as forwarding nodes, offering routes to other destinations, it is expected that such nodes actually forward data packets once the route is setup. However , malicious nodes deviate from this expected behaviour and maliciously drop data packets[11] thereby disrupting transmission. Such Malicious nodes are identified by installing a watchdog and a PathSelector in the Ad-hoc network on every server.

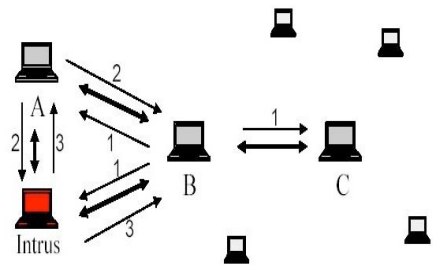

## VIII.WATCHDOG

The watchdog identifies misbehaving nodes, while the Pathselector avoids routing packets through these nodes. When a node forwards a packet, the watchdog verifies that the node forwards the packet to all its neighbours. If the node does not forward the packet even to one of its neighbour,then it is misbehaving. The PathSelector[12] uses this knowledge of misbehaving nodes to choose the network path that is most likely to deliver packets.

The watchdog is implemented on every server by maintaining a buffer of neighbours for each and every node. When a node transmits packet to its neighbour,the corresponding entry in the buffer is forgotten by the watchdog, since it has been forwarded on. If an entry for a corresponding node's neighbour has remained in the buffer for longer than a certain timeout, the watchdog increments a failure rate for the node and determines that the node is misbehaving.

## IX.PATH SELECTOR

The watchdog, the pathselector is run by each server. Each Server maintains a rating for every other node it knows about in the VHR. Path selector works during the contention scheme. If a node is misbehaving, the watchdog will identify such a node before route discovery itself and assigns a failure rate to such a node. The Pathselector then compares the rating for the nodes that win the contention scheme with this failure rate to determine whether they are misbehaving. If so, they are not included in the route at all. Thus they are avoided from becoming part of the route.

*An Example:*

Suppose that a network is consisting of the nodes labeled S(source), D (destination) and from alphabet A to I. The source wishes to communicate with the destination. At first, the source select the mode as 1 based on the required security level of application.

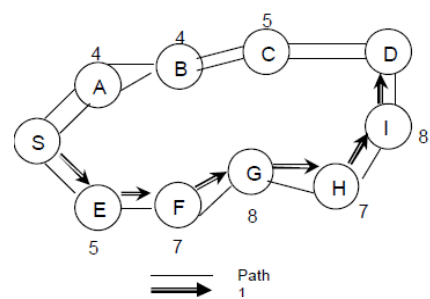

| Node ID | Trusted Node | Trusted Route |
|---------|--------------|----------------|
| A | 4 | Path Selection |
| E | 5 | Path Selection |

Selection of Source Node & Path Modification

The numbers shown closer to each node indicate their corresponding trust level. Node S to communicate with node D broadcasts rreq to it's neighbors A and E. There are two possible paths from node S to D: S-A-B-C-D (path1), S-E-F-G-H-I (path2). Node A tries to authenticate the source node S. It checks it's trust table. If S is trusted, A accepts the rreq message, update the location field and TUSN in it's neighbor table and broadcast the rreq to the next node. If S cannot be trusted, A drops the rreq. If S is not in A's table, A send a trust_request to S. If the response is 'yes', A stores the information in it's trust table and rebroadcasts the rreq. When the response is not received within a limited time, node A drops the rreq. As a result node A forwards to B, B forwards to C and C forwards to destination D. Similarly in path 2, E forwards to F, F forwards to G, G forwards to H, H forwards to I and I to destination D.

The destination D unicasts[13] the rrep to C and I separately. Node C send the reply to node B. Node B forward the packet to A. But before sending, each node attaches the trust level of the node from where it just received the rrep. Upon receiving the rrep, each node update the recent destination list. The node attaches the trust level of C to trust string. So the trust string now contains the value 5. Node B forwards the rrep to A. Now the value of trust string is 54. The process continues until it reaches the source node. So the final value of trust string for the path 1 is 544. Similarly in path 2 node I forwards the rrep to I. The process will be similar as in path1. The final value of trust string for the path 2 is 87875.

Now the source waits for a predefined time period to select the best route. The application requires trusted path for communication. The average trust weight of path 1 is 4.33 and trust weight of path 2 is 7. Hence path 2 is selected.

## X. SYBIL ATTACKS

The contention scheme proposed in the AO2P Protocol is vulnerable to malicious nodes that try to become part of the route by claiming False Identities for themselves. Their False Identity is such that, that the malicious node becomes the best choice for data transmission. Such nodes are called as SYBIL NODES. In order to identify Sybil nodes the following method is proposed.
When a node sends the hrep packet to the sender, the sender forwards the packet to the server for validation. The server then sends a message to the location specified in the Pseudo ID of the hrep packet forward. If the node responds to it, then the server knows that the node is legitimate and that its position is geographically correct. If the node does not respond to the message, then the server knows that the node is a Sybil node and that its malicious. Therefore it alerts the sender. The sender then ignores this hrep packet and chooses the next best choice.

## XI. DESTINATION MOBILITY

Once a route is found between the source and the destination, the destination responds to the source by sending the rrep message. There is a possibility that by the time rrep message reaches the source, the destination might have moved to the new location. Hence this when unnoticed can lead to large position errors. In order to keep the source informed about the destination's mobility, the destination keeps sending the alert message to its previous hop telling that it has changed its position and any reference to it for data packet forwarding be informed to the VHR server. This alert message is forwarded in the reverse path until it reaches the source that initiated the transmission. The VHR server when intimated will know about the destination's new location and hence the data packets can be sent to the destination properly.

## XII. CONCLUSION

The protocol follows different routing mechanism based on the security level required by application. In mode 1, the packets are routed along the trusted path based on the trust factor of the nodes. In mode2, the packets are routed along the shortest path based on hop count. The protocol uses a mechanism to detect and overcome the effect of falsified position information in geographic routing position. The protected position information reduces the routing overhead and increase the security of routing. Destination position alone is revealed for routing purposes.

## Authors

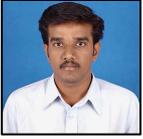

Sudhakar Sengan is a faculty member in Department of Computer Science and Engineering , Nandha College of Technology , Erode , TamilNadu, India. He received his M.E.(CSE) in 2007 from Anna University , Chennai , TamilNadu, India.And Pursing Ph.D in Anna University–Coimbatore, Chennai, TamilNadu,India.His areas of research interests include Mobile Computing and Network Security. He is Published National Conferences 4 and International Conferences 2.

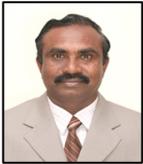

DR. S. Chenthur Pandian Ph.D is the Principal in Selvam College of Technology, Namakkal , Tamil Nadu, India.He received his B.Sc.(Maths) from Madurai Kamaraj University, Madurai in 1983 and AMIE from Institute of Engineers (India), Calcutta in 1984 and M.E. from Punjab University, Chandigarh in 1994 and Ph.D. in Fuzzy Application for Power Systems from Periyar Univeristy, Salem in 2005. He published National Journals 3 , International Journals 4,Conferences 34.And he published around 3 books. He is a member of MIEEE(USA), MISTE. His research interests include Fuzzy Application for Power Systems, Power Electronics, Neuro-Fuzzy System.